\documentclass[aps,prb,twocolumn,amsmath,amssymb,footinbib,superscriptaddress,notitlepage]{revtex4-2}
\usepackage{graphicx}
\usepackage{amsmath,amssymb}
\usepackage{dcolumn} 
\usepackage{bm} 
\usepackage{xcolor}
\usepackage[utf8]{inputenc}
\usepackage{appendix}
\usepackage{listings}
\usepackage{float}
\usepackage{epstopdf}
\usepackage{hyperref}
\usepackage[all]{hypcap}
\usepackage{geometry}
\usepackage{natbib}
\usepackage{soul}

\begin{document}

\title{Quantum Sensing of Single Phonons via Phonon Drag in Two-Dimensional Materials}
\author{Ali Kefayati}
\affiliation{Department of Electrical Engineering, University at Buffalo, The State University of New York, Buffalo, New York 14260, USA.}
\author{Jonathan P. Bird}
\affiliation{Department of Electrical Engineering, University at Buffalo, The State University of New York, Buffalo, New York 14260, USA.}
\author{Vasili Perebeinos}
\email{vasilipe@buffalo.edu}
\affiliation{Department of Electrical Engineering, University at Buffalo, The State University of New York, Buffalo, New York 14260, USA.}

\date{\today}

\begin{abstract}

The capacity to electrically detect phonons, ultimately at the single-phonon limit, is a key requirement for many schemes for phonon-based quantum computing, so-called quantum phononics. Here, we predict that by exploiting the strong coupling of their electrons to surface-polar phonons, van der Waals heterostructures can offer a suitable platform for phonon sensing, capable of resolving energy transfer at the single-phonon level. The geometry we consider is one in which a drag momentum is exerted on electrons in a graphene layer, by a single out-of-equilibrium phonon in a dielectric layer of hexagonal boron nitride, giving rise to a measurable induced voltage ($V_{\rm drag}$).
Our numerical solution of the Boltzmann Transport Equation shows that this drag voltage can reach a level of a few hundred microvolts per phonon, well above experimental detection limits. Furthermore, we predict that $V_{\rm drag}$ should be highly insensitive to the mobility of carriers in the graphene layer and to increasing the temperature to at least 300 K, offering the potential of a versatile material platform for single-phonon sensing.

\end{abstract}

\maketitle

\begin{section}{Introduction}

A key requirement of quantum computing is the on-demand generation and coherent manipulation of quantum states. From a practical perspective, the ability to implement these operations using chip-scale integrated circuits is highly desired, as it would open up the capability to fully leverage the many advantages of the mature microelectronics industry. However, a fundamental barrier to solid-state quantum computing has long been understood to arise from the role of phonons. At nonzero temperatures, these bosonic modes can be occupied with broad distributions of energy and momentum, allowing them to function as an efficient “bath” whose many degrees of freedom can efficiently randomize (or decohere) quantum information. Although the manner in which a densely populated phonon bath destroys quantum coherence has long been understood, it has only recently become appreciated that phonons may instead provide an effective means of transmitting quantum information when excited coherently in sufficiently small numbers~\cite{chu2018creation,gustafsson2014propagating,manenti2017circuit,noguchi2017qubit,bolgar2018quantum, moores2018cavity, satzinger2018quantum}. The strong coupling of phonons to other quasiparticles (especially electrons or photons) makes them well suited to this task. Furthermore, the physical patterning of bulk crystals can be exploited to implement phononic crystals ~\cite{schuetz2015universal, kurizki2015quantum, reinke2016phonon}, in strong analogy with their photonic counterparts, or to realize resonant structures that can be selectively coupled to single quanta ~\cite{poot2012mechanical,barfuss2015strong}. Various schemes for phonon-mediated quantum transduction have been proposed~\cite{Safavi_Naeini_2011,lemonde2018phonon,weber2010quantum,chen2018orbital,chen2019engineering,vermersch2017quantum,Pollanen2022}, and phonons have also been suggested as a means of mediating quantum entanglement ~\cite{hucul2015modular}. In other work, the development of phononic circuits for applications in quantum sensing and signal processing has been  emphasized~\cite{Aref2016,safavi2019controlling}. Stark shift measurements as a function of the number of phonons~\cite{Leak_NCom2017} have been demonstrated and are based on the real part of the electron-phonon self-energy. Other phonon detection schemes use optomechanical response~\cite{Li2007}, including the famous LIGO gravitational wave interferometer~\cite{LIGO}. In contrast to these different approaches, a phonon-drag detection scheme is proposed that utilizes a 2D material platform and which takes advantage of the imaginary part of the self-energy, leading to high detection efficiency and simplicity of implementation.

For phononic control to approach the levels of sophistication that have already been achieved for electrons and photons, there are a number of critical issues that need to be addressed. Key among these is the need to detect phonons in real-time in an electrical measurement, using approaches that can ultimately be scaled to the limit of single-phonon resolution. In this work, we propose and predict the quantitative performance of a single-phonon detector that is implemented in a heterostructure of monolayer (or AB-stacked bilayer) graphene and multilayer hexagonal boron nitride (hBN). It has been known for four decades that remote phonon scattering hinders the mobility of semiconductors grown on polar substrates, in which ionic motion generates an electric field that extends into the semiconductor~\cite{Hess1979,fischetti2001effective}. In this work, we propose to make use of remote phonon scattering to implement phonon detectors. Two-dimensional materials such as graphene have recently attracted increasing interest for application in quantum computing technology~\cite{liu20192d}. When choosing the particular geometry of Fig.~\ref{Fig:Diagram}a, we are motivated by the fact that hBN is a notable phononic material, having branches of phonon polariton (near 100 meV and over the range from ~175
– 200 meV) that exhibit a hyperbolic character due to the strong optical anisotropy. This allows thin slabs of this material to function as efficient waveguides for propagating phonons ~\cite{dai2015graphene,principi2017super,Low2017}, a characteristic that has been exploited ~\cite{yang2018graphene} to achieve rapid cooling of the hot carrier in graphene / hBN-based transistors. At the same time, the presence of the tunable, high-conductivity electron gas in graphene renders it well suited for the drag-based detection of phonons in the hBN. To demonstrate this, we consider a situation in which a single surface polar phonon (SPP) is excited in the hBN and travels along the heterostructure while interacting with electrons in the graphene layer. This process leads to the development of the drag-based voltage shown in Fig.~\ref{Fig:Diagram}a.



In a heterostructure formed between graphene and monolayer hBN, the inherent 2D nature of the materials gives rise to hybrid plasmon-phonon polaritons that propagate parallel to the plane of their interface. However, the situation is very different when graphene is deposited in thicker layers of hBN (in the range of $>$1 – 100 nm), in which the phonon polaritons exhibit a hyperbolic character, capable of propagation with large energy and momentum losses. Optical phonons injected into hBN will initially propagate with a ray-like character before decaying via crystal anharmonicity over a characteristic distance of a few tens of nanometers, allowing the formation of long-lived~\cite{Basov_lifetime_2018,Basov_lifetime_2021} SPPs at the boundary between the hBN and graphene layers. Our calculated results demonstrate that the drag voltage that develops in such structures is on the scale of a few hundred microvolts for a device one micrometer wide, well above the detection limit in typical experimental setups.

\end{section}

\begin{section}{Modeling approach}

The interplay of charge carrier flow with phonon transport (and vice versa) has a long history of discussion in thermoelectrics ~\cite{bailyn1958transport, dugdale1967anisotropy, jonson1980mott}. At the same time, interest in the problem of phonon drag has been revived in the context of low-dimensional materials ~\cite{scarola2002phonon, koniakhin2013phonon, kubakaddi2010enhancement, ghawri2020excess, waissman2021measurement, yalamarthy2019significant}. We assume a quasiparticle picture of electrons and phonons in our study, employing the Boltzmann transport formalism in our calculations~\cite{ziman2001electrons}:
\begin{eqnarray}
     \frac{e\Vec{F}}{\hbar} \frac{\partial f_{k}}{\partial \Vec{k}} = \left(\frac{\partial f_{k}}{\partial t} \right)_{e-ph} + \left(\frac{\partial f_{k}}{\partial t} \right)_{imp} .
\label{Equ:BTE}
\end{eqnarray}
Here, $f_{k}$ is the distribution function of electrons in momentum space, $k$ is the electron wavevector, including the band index, $\Vec{F}$ is the external electric field that acts as the driving force, $\Vec{v}_{k}$ is the group velocity of electrons and $\hbar$ is the Planck constant. The terms on the right-hand side of the equation describe collision integrals due to electron-phonon ($e-ph$) and electron-impurity ($imp$) scattering. 
\begin{figure*}
  \centering
 \includegraphics[width=1.0\textwidth]{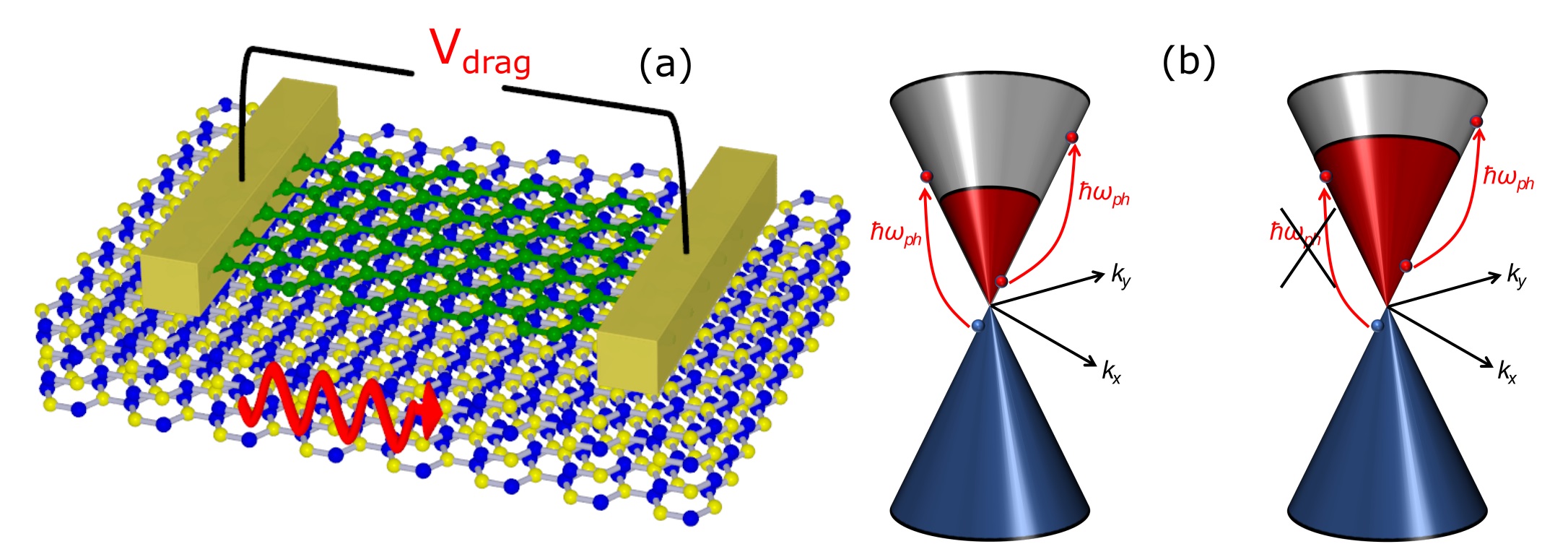}
 \caption{(a) Schematic representation of a stacked heterostructure of graphene (green bonds) and hBN (grey bonds) that exploits the phonon-drag effect for phonon (red arrow) detection. (b) Phase space for electron excitation via e-SPP interaction in monolayer graphene. When $E_F<\hbar\omega_{ph}$, there are two possible kinds of excitation, inter-band (valence to conduction band) and intra-band (within the conduction band). When $E_F>\hbar\omega_{ph}$, the inter-band transition is no longer allowed.}
\label{Fig:Diagram}
\end{figure*}

 \begin{figure*}
  \centering
 \includegraphics[width=1.0\textwidth]{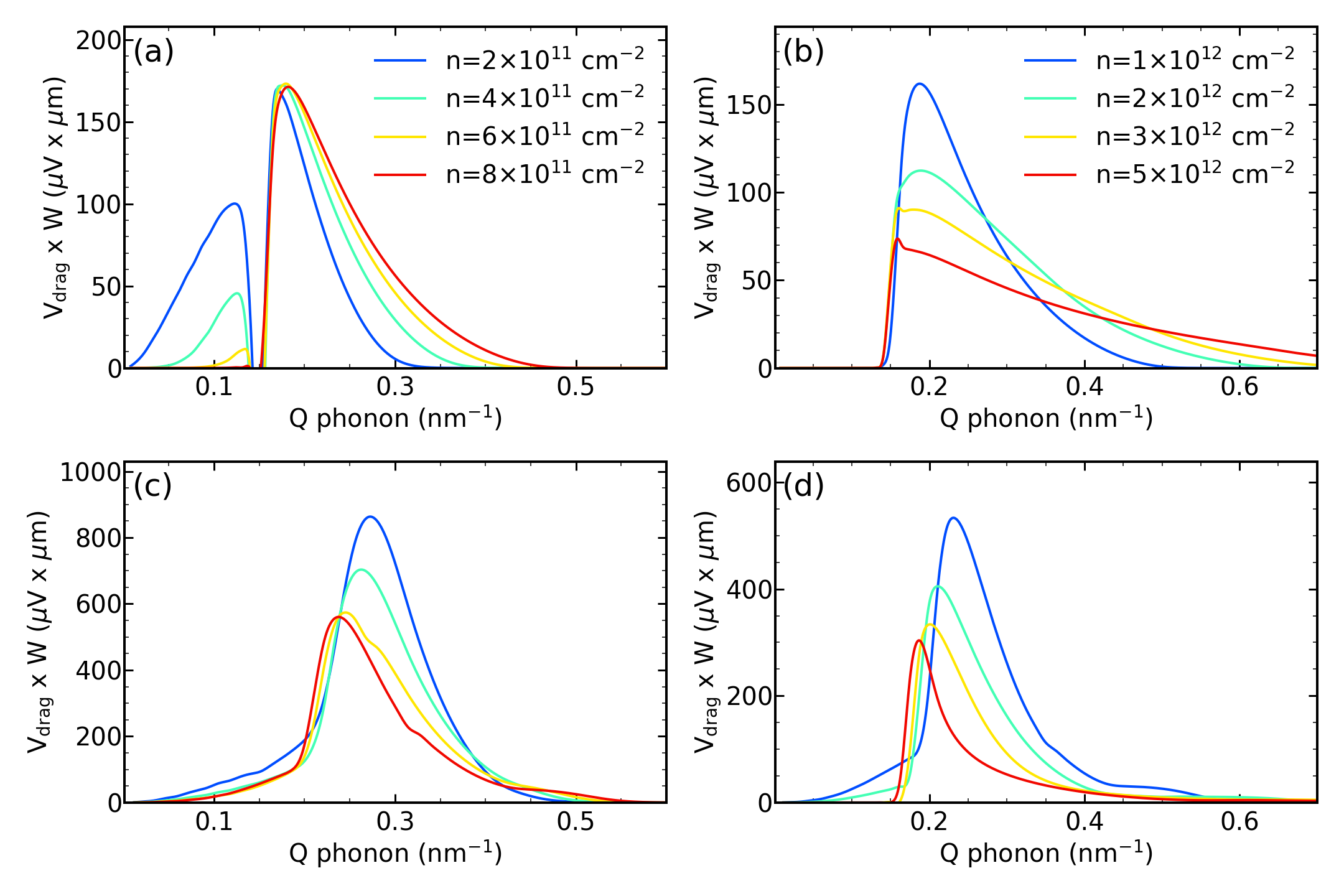}
 \caption{Phonon drag voltage $V_{\rm drag}$ as a function of the hBN SPP wavevector, and for various carrier densities in the graphene layer. Panels (a) and (b) show results for monolayer graphene, while panels (c) and (d) are for bilayer graphene. The legend of panel (a) applies to panel (c) also, while that of panel (b) applies to panel (d). The calculations assume $T=50$ K and $t_{\rm hBN}=\infty$.}
\label{Fig:FigI}
\end{figure*}

The collision integral for electron-phonon scattering is given by:
\begin{widetext}
\begin{equation}
\begin{split}
 \left(\frac{\partial f_{k}}{\partial t} \right)_{e-ph} &= -\sum_q{ \vert M_{\mathbf{k}\mathbf{q}}\vert^2}
\bigg [ \big [f_k(1-f_{k-q})(1+n_q) - f_{k-q}(1-f_k)n_q \big] \delta\big(E_k-E_{k-q}-\hbar\omega_q\big)\\
  &+\big[f_k(1-f_{k+q})n_{q} - f_{k+q}(1-f_k)(1+n_{q}) \big]  \delta\big(E_k-E_{k+q}+\hbar\omega_{q}\big) \bigg ] . 
  \label{Equ:e-ph}
\end{split}    
\end{equation}
\end{widetext}
In this equation, $n_q$ is the number of occupied phonon modes with momentum $q$ and energy $\hbar\omega_q$, $E_k$ is the electron energy, and the $\delta$-function ensures energy conservation. $\vert M_{\mathbf{k}\mathbf{q}}\vert^2$ is the coupling constant of the e-ph interaction. The details of the electron-SPP scattering are given in Appendix~\ref{sec:level1} and intrinsic electron-phonon interactions are taken into account following Ref.~\cite{screenSPP_PRL2022}. 

The collision integral for Coulomb impurity scattering is added using a standard procedure~\cite{Hwang2008,DasSarma2010,Min2010,DasSarma_2011}. Unless stated otherwise, we choose the impurity concentration to give a typical mobility of $\sim1000$ cm$^2$/Vs for the graphene devices. 

The problem we consider is one in which the phonon systems of both materials are initially in thermal equilibrium and in which we then assume that a single SPP (of wavevector $Q$ in direction $\alpha$) is excited in the hBN layer. As a result of this, the distribution function in Eq.~(\ref{Equ:BTE}) changes by an amount $\Delta f_{k}$, leading to an excess current density $\Delta j_{\rm drag}$. We define the resulting drag voltage $V_{\rm drag}$ via:
\begin{eqnarray}\label{eq:Vdrag}
   &&\Delta j_{\rm drag}=\frac{e\sum_{k} \Delta f_{k}v_{\alpha k} }{WL}=\sigma\frac{V_{\rm drag}}{L}
   \nonumber \\ 
   &&V_{\rm drag}=\frac{e\sum_{k} \Delta f_{k}v_{\alpha k} }{\sigma W},
\end{eqnarray}
where $\sigma$ is the electrical conductivity and $W$ and $L$ are the width and length of the device, respectively. In simulations, we use a supercell approach such that the $k$-point mesh $N_x\times N_y$ of the Brillouin zone defines the corresponding area of the device in real space, i.e., $W \times L= N_x\times N_y \times A_{c}$, where $A_c$ is the area of the primitive unit cell.
According to  Eq.~(\ref{eq:Vdrag}), the drag voltage is inversely proportional to the width of the conducting graphene channel. In the analysis that follows, we therefore report the results of the product $V_{\rm drag}\times W$.  It should be noted that our choice to characterize the responsivity of the phonon sensor in terms of the drag voltage $V_{\rm drag}$ leads to a counterintuitive conclusion that the responsivity is independent of the quality of the graphene sample and of the strength of the impurity scattering (see discussion below). This result is a consequence of the form of Eq.~(\ref{eq:Vdrag}), in which the $\Delta f_{k}$ caused by an out-of-equilibrium phonon is proportional to the electrical conductivity, leading to $V_{\rm drag}$ being nearly independent of mobility. The phonon-induced current is certainly sensitive to the quality of graphene, however.

\end{section}

\begin{section}{Results and Discussion}

Through a process of absorption, the out-of-equilibrium SPP excites an initial electron into a higher energy state (see Fig.~\ref{Fig:Diagram}). The conservation of energy and momentum requirements impose a dependence of $V_{\rm drag}$ on the SPP wavevector and the Fermi energy $E_F$ in the graphene. Fig.~\ref{Fig:Diagram} illustrates two possible electron excitations that can arise from absorption of the out-of-equilibrium phonon: an inter-band transition of an electron from the valence- to the conduction-band, as shown in Fig.~\ref{Fig:Diagram}a, or an intra-band transition between two states in the conduction-band, as shown in Figs.~\ref{Fig:Diagram}a and~\ref{Fig:Diagram}b. Energy conservation prohibits interband transitions once the Fermi level is larger than the SPP energy \emph{i.e.,} $E_F>\hbar\omega_{ph}$ in monolayer or bilayer graphene.

\begin{figure*}[htp]
  \centering
 \includegraphics[width=1.0\textwidth]{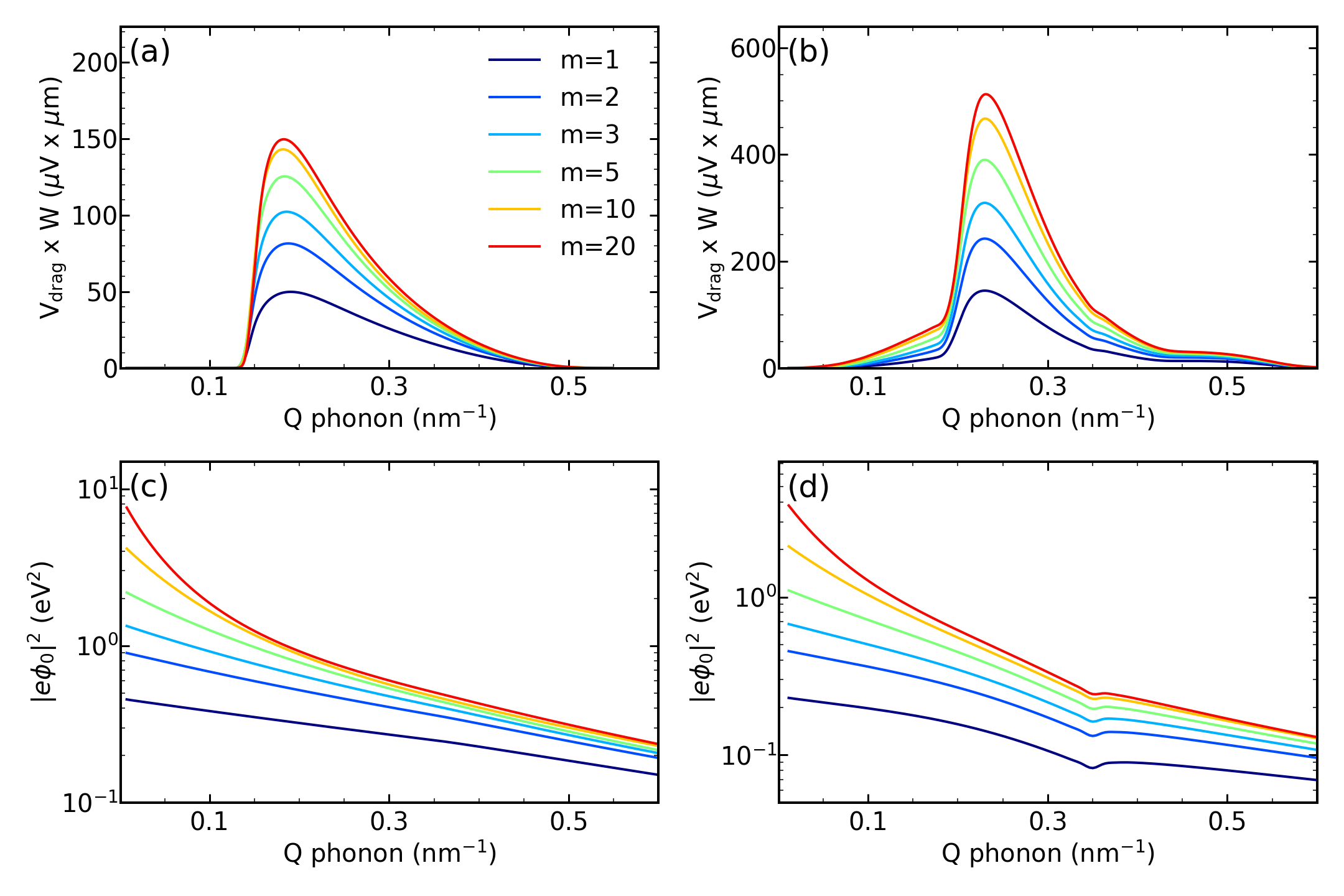}
 \caption{Phonon drag voltage $V_{\rm drag}$ as a function of hBN SPP wavevector in (a) monolayer and (b) bilayer graphene. Results are shown for different hBN layer number $m$, or thickness $t_{\rm hBN}=m\times 3.4$ \AA. Panels (c) and (d) depict the e-SPP scattering potential from Eq.~(\ref{eq:e-SPP}) for monolayer and bilayer graphene, respectively. The weak kink at $Q=0.34$ nm$^{-1}$ in panel (d) is due to the polarization function anomaly at $Q=2k_F$~\cite{Hwang2008,DasSarma2010}. The calculations assume $T=50$ K and $n=10^{12}$ cm$^{-2}$.}
\label{Fig:FigII}
\end{figure*}

\begin{figure}[htp]
  \centering
 \includegraphics[width=0.48\textwidth]{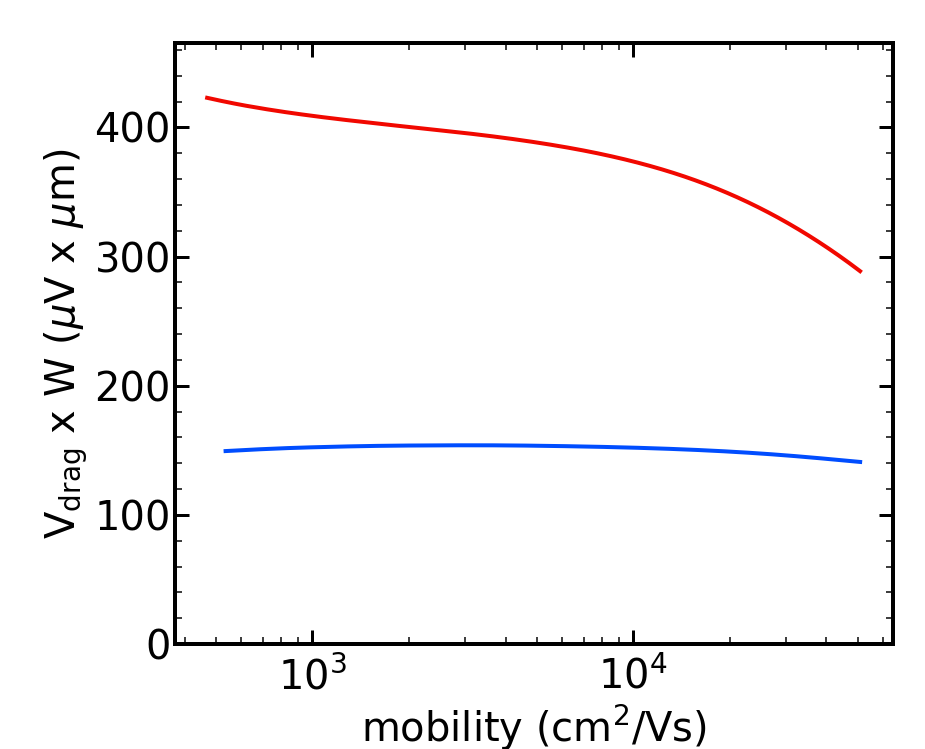}
 \caption{Phonon drag voltage $V_{\rm drag}$ in monolayer and bilayer graphene as a function of the Coulomb-scattering limited mobility, which is controlled by the impurity concentration. The fixed parameters here are: $T=50$ K, $n=10^{12}$ cm$^{-2}$, $t_{\rm hBN}=\infty$, and $Q=0.20$ nm$^{-1}$/0.27 nm$^{-1}$ in monolayer/bilayer graphene.}
\label{Fig:FigIII}
\end{figure}

\begin{figure}
  \centering
 \includegraphics[width=0.48\textwidth]{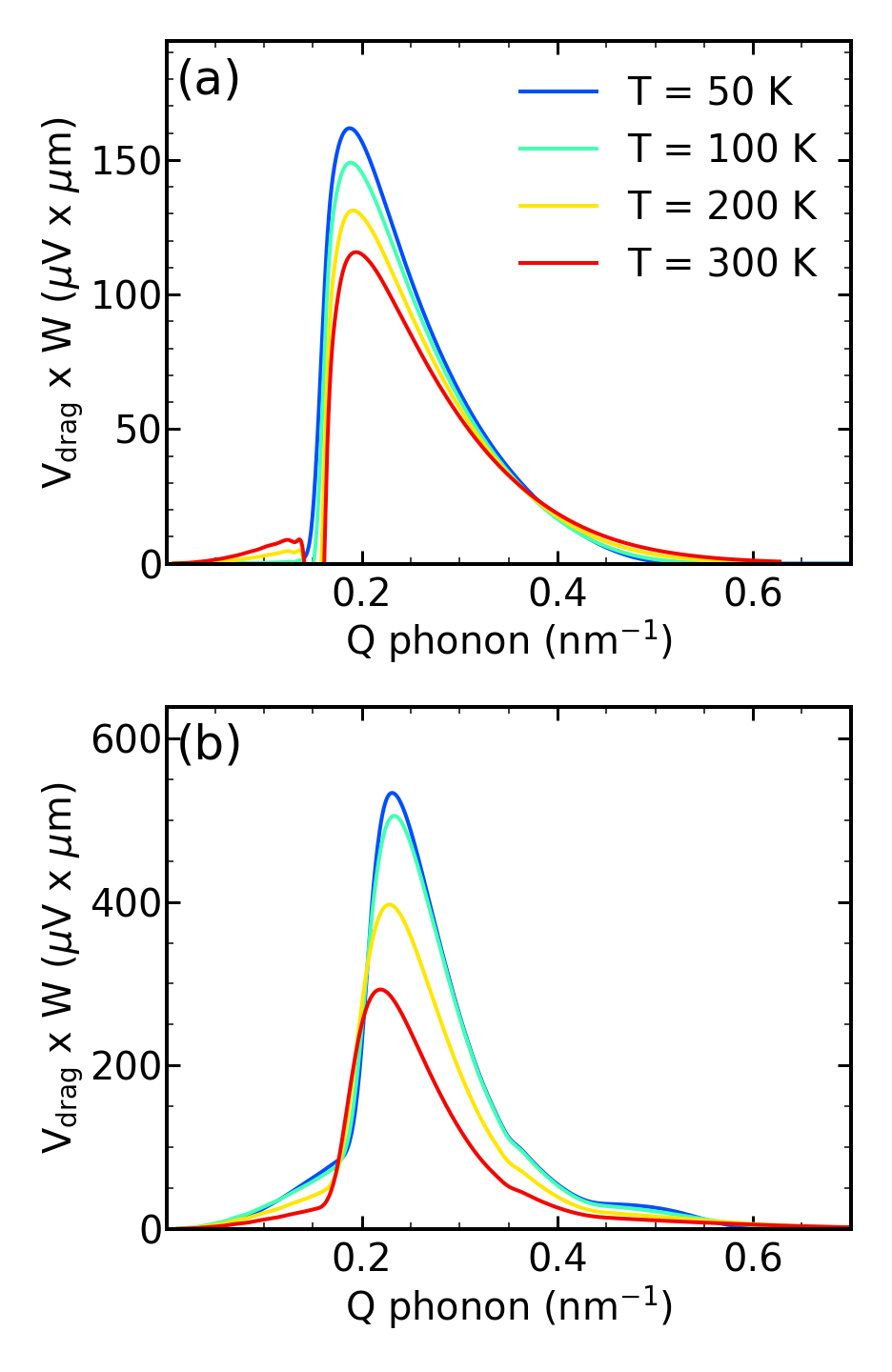}
 \caption{Phonon drag voltage $V_{\rm drag}$ as a function of hBN SPP wavevector in (a) monolayer (blue) and (b) bilayer (red) graphene, for four different temperatures (indicated in panel (a)). The calculations assume $n=10^{12}$ cm$^{-2}$ and $t_{\rm hBN}=\infty$.}
\label{Fig:FigIV}
\end{figure}

 In Fig.~\ref{Fig:FigI} we plot the variation of $V_{\rm drag}$ as a function of the SPP wavevector, for various carrier densities in both the monolayer (Figs.~\ref{Fig:FigI}a and b) and bilayer graphene (Figs.~\ref{Fig:FigI}c and d). In the monolayer case, the drag signal exhibits two distinct peaks, the first of which, at small wavevector, is associated with inter-band excitation of electrons from the valence- to the conduction-band.  The wavevectors at which the peaks in the drag signal occur reflect the details of the phase space for scattering, which is goverened by conservation of energy and momentum and the Pauli blocking principle. The amplitude of these peaks is determined by the strength of the electron-phonon matrix elements. The finite temperature primarily introduces thermal smearing of the electronic states. As indicated by the schematic of Fig.~\ref{Fig:Diagram}, these transitions are able to satisfy energy and momentum conservation laws at small SPP wavevector. They are cut-off at wavevectors greater than $Q_{c1}=\omega_{\rm ph}/v_{\rm F}\approx 0.15$ nm$^{-1}$, where $v_{\rm F}\approx10^6$ m/s is the Fermi velocity in the monolayer and $\hbar\omega_{\rm ph}\approx 100$ meV is the phonon energy, see Appendix~\ref{sec:level1}. The inter-band transitions are to be contrasted with those responsible for the second peak in the drag signal, seen at larger wavevectors, which instead involve intra-band processes.

 Turning next to the results for bilayer graphene (Figs.~\ref{Fig:FigI}c and~\ref{Fig:FigI}d), we see that the small-wavevector peak found in the monolayer case is replaced now by a shoulder-like feature for wavevectors $Q<Q_{c2}$, where $Q_{c2}=(2m\omega_{\rm ph}/\hbar)^{1/2}\approx 0.3$ nm$^{-1}$, $m\approx0.035\times m_e$ is effective mass in bilayer graphene~\cite{massexp2011}, and  $m_e$ is free-electron mass. The absence of the inter-band drag peak at small wavevectors is a consequence of the reduced phase space available for electron scattering in the parabolic bands of the bilayer. One can estimate the carrier density at which the feature in the drag signal due to inter-band transitions should be suppressed, in monolayer or bilayer graphene, as $n_{c1}=Q_{c1}^2/\pi\approx7.3\times10^{11}$ cm$^{-2}$ and $n_{c2}=Q_{c2}^2/\pi\approx2.9\times10^{12}$ cm$^{-2}$, respectively. These estimates are consistent with the trends indicated in Fig.~\ref{Fig:FigI}.
 
 As doping increases in monolayer graphene, the wavevector range over which phonon detection can be achieved increases, as reflected by the increase in the width of the peak of the higher moment in Fig.~\ref{Fig:FigI}b. Similar phase-space arguments apply to bilayer graphene. One can estimate a wavevector cut-off for the intra-band transitions according to $Q_{\rm max1}\approx 2 k_F+ \frac{\omega}{v_{\rm F}}$ in monolayer graphene and $Q_{\rm max2}\approx 2 k_F+ \frac{m\omega}{\hbar k_{\rm F}}$ in bilayer graphene, where $k_F\approx\sqrt{\pi n}$ is the Fermi wavevector and we have used an effective Fermi velocity $\hbar k_{\rm F}/m$ in bilayer graphene. The largest carrier density of $5\times10^{12}$ cm$^{-2}$ used in Figs.~\ref{Fig:FigI}b and~\ref{Fig:FigI}d translates to corresponding estimates of $Q_{\rm max 1,2}\approx$ 0.95 nm$^{-1}$. However, the detection signal dies at smaller wavevectors of about $0.6$ nm$^{-1}$ in the case of bilayer graphene, which is due to the decay of the electron-SPP matrix element at large wavevectors, as discussed in Appendix~\ref{sec:level1} and illustrated in Figs.~\ref{Fig:FigII}c and~\ref{Fig:FigII}d.

To provide insight into the dependence of the drag signal on the thickness of the hBN, we calculated $V_{\rm drag}$ (at a fixed carrier density $n=10^{12}$ cm$^{-2}$) for several values of $t_{\rm hBN}$. The results are shown in Figs.~\ref{Fig:FigII}a and~\ref{Fig:FigII}b, for monolayer and bilayer graphene, respectively. As the number of layers is reduced, the electric field associated with the SPP also decreases. However, this variation is relatively weak, and $V_{\rm drag}$ is reduced by only a factor of three relative to the bulk case, by the time that the single layer limit is reached. Conversely, as the hBN thickness increases from the single-layer limit, the drag signal saturates once the number of hBN layers reaches around twenty. 

To better understand the $Q$-vector dependence of the drag voltage at large momenta, we note that the electron-SPP coupling is Coulombic in nature. According to  Eq.~(\ref{eq:e-SPP}), the Fourier transform of this coupling has a strong momentum dependence. In Figs.~\ref{Fig:FigII}c and~\ref{Fig:FigII}d, we plot the scaling of the electron-SPP scattering potential as a function of the phonon wavevector to vary the number of hBN layers. The momentum dependence of $V_{\rm drag}$ reported here is a consequence of the product of the electron-SPP coupling and the phase space available for electron excitation due to the non-equilibrium phonon.

The trend apparent in Figs.~\ref{Fig:FigI}b and~\ref{Fig:FigI}d, for the amplitude of the higher moment peak to decrease with increasing carrier density, can be attributed to the dependence of electrical conductivity on density, \emph{i.e.}, $\sigma=en\mu=e^2n\tau/m$, where $\mu$ is the carrier mobility and $\tau$ is the effective scattering time. According to Eq.~(\ref{eq:Vdrag}), $V_{\rm drag}$ is inversely proportional to $\sigma$, which accounts for the reduction mentioned above of $V_{\rm drag}$ with increasing density. At the same time, one should keep in mind that the increase in the distribution function ($\Delta f_k$) due to the excitation of the single out-of-equilibrium phonon is proportional to $\tau$. This means that the influence of impurity scattering on the drag voltage should be relatively weak. In fact, in Fig.~\ref{Fig:FigIII} we show that $V_{\rm drag}$ is almost independent of $\mu$, whose value is largely determined by the concentration of impurities.  

We have also explored the dependence of the drag signal on the equilibrium lattice temperature. The influence of this parameter is illustrated in Fig.~\ref{Fig:FigIV}, for both monolayer and bilayer graphene. (The calculations are performed for a representative density $n=10^{12}$ cm$^{-2}$, and for $t_{\rm hBN}=\infty$). Due to the large SPP energy ($\sim$100 meV), 
$V_{\rm drag}$ is reduced by only about 30\% in monolayer graphene (and 50\% in bilayer graphene), when the 
temperature increases from 50 to 300 K. This robust character opens up the possibility of realizing single-phonon detectors that are capable of functioning at room temperature.  

When considering schemes for single-phonon detection, it is important to keep in mind the fact that phonons have a finite lifetime due to phonon-phonon decay, and a transit time that is determined by the transistor size and the phonon velocity $v_{\rm ph}$. For a field-effect transistor with channel length $L\sim 1 \mu$m, the transient time can be estimated as $\tau_{\rm tr}=L/v_{\rm ph}\sim1$ ns, setting an upper bound for the limited lifetime of anahrmonic decay $\tau_{\rm ph}$. Transient effects can therefore be modeled by the Boltzmann transport equation as follows: 
\begin{eqnarray}
     \frac{\partial\Delta f_{k}(t)}{\partial t} = \frac{\Delta f_{k}(t)}{\tau} + \exp{\left(-\frac{t}{\tau_{ph}}\right)}\left(\frac{\partial f_{k}}{\partial t} \right)_{\rm NQ}
\label{Equ:BTE_time}
\end{eqnarray}
where, without loss of generality, we have used the relaxation time approximation to describe electron-impurity scattering in terms of an effective scattering time $\tau$. The NQ subscript for the collision integral denotes the contribution to electron scattering due to the single out-of-equilibrium phonon. The solution of Eq.~(\ref{Equ:BTE_time}) is given by: 
\begin{eqnarray}
     \Delta f_{k}(t)= \Delta f_{k}\frac{\tau_{\rm ph}}{\tau_{\rm ph}+\tau} \exp{\left(-\frac{t}{\tau_{\rm ph}}\right)},
\label{Equ:BTE_time}
\end{eqnarray}
where $\Delta f_{k}$ is the steady-state solution of Eq.~(\ref{Equ:BTE}). According to Eq.~(\ref{eq:Vdrag}), the drag voltage will be reduced by a factor of $\tau_{\rm ph}/(\tau_{\rm ph}+\tau)$ and will have the same exponential time dependence as appears in Eq. ~(\ref{Equ:BTE_time}). The steady-state solution for the drag voltage corresponds to the limit $\tau_{ph}\rightarrow\infty$ in the transient solution:  
\begin{eqnarray}\label{eq:Vdrag_time}
V_{\rm drag}(t)=V_{\rm drag}\frac{\tau_{\rm ph}}{\tau_{\rm ph}+\tau}\exp{\left(-\frac{t}{\tau_{\rm ph}}\right)}
\end{eqnarray}
where $V_{\rm drag}$ is given by Eq.~(\ref{eq:Vdrag}) and reported in Figs.~\ref{Fig:FigI}-\ref{Fig:FigIV}. While counter-intuitive, Eq.~(\ref{eq:Vdrag_time}) suggests that lower-mobility graphene, with shorter scattering time, should perform better in measuring transient phonon-drag voltage signals.

An important parameter for sensor characterization is the noise equivalent power (NEP), which characterizes the signal-to-noise ratio of the phonon detectors. 
Phonons follow Bose-Einstein statistics, meaning that the noise considerations relevant to single-photon detectors should also be applicable to the single-phonon detector proposed here. Following Ref.~\cite{bolometer1994}, we can estimate the NEP due to the Johnson-Nyquist noise contribution as a ratio of Johnson noise $\sqrt{4k_bTR}$ and voltage responsivity, where $R= L/(W e  n  \mu)$ is the resistance of the device. The voltage responsivity can be estimated as $V_{\rm drag}/(\hbar\omega_{\rm ph}/\tau_{\rm tr})$, where the transient time $\tau_{tr}\sim1$ ns. Using realistic device parameters: $W=L=1$ micron, $n=10^{12}$ cm$^{-2}$, $\mu=1000$ cm$^2$/Vs, and a typical $V_{drag}=100$ $\mu$V, we thus estimate the NEP to be ~0.16 fWHz$^{-1/2}$.

Fluctuations in the phonon distribution are not taken into account in our Boltzmann Transport approach, and any temperature dependencies arises from the thermal smearing of the electron distribution function. Fluctuations in the phonon population are known to cause noise in conductivity~\cite{Jindal1982}. We emphasize that here we consider phonon drag due to a single SPP phonon in h-BN, a mode whose thermal population is very small at room temperatures.

\end{section}

\begin{section}{Conclusions}

In conclusion, we have shown theoretically that the phonon-drag effect in two-dimensional layered materials can enable the sensitive, quantum detection of phonons, with the potential of operation down to the single-phonon level. The strong coupling between electrons and SPPs in the coupled conductive and dielectric layers is predicted to give rise to a drag voltage as large as a few hundred microvolts per phonon, well above the experimental detection limit. The drag voltage is, moreover, predicted to be relatively insensitive to variation in the mobility of the graphene layer and should exhibit only a weak temperature dependence. These characteristics relax the need for high-mobility detectors that operate at ultralow temperatures. By varying the Fermi level in graphene using a suitable gate, the phonon detectors described here should act as efficient energy-resolved phonon sensors, since momentum conservation laws set stringent requirements on the range of phonon wavevectors that can be detected. Our study provides further evidence of the outstanding potential of 2D heterostructures for use in quantum information science and quantum sensing~\cite{liu20192d}.

\end{section}

\begin{acknowledgements}
This material is based upon work supported by the Air Force Office of Scientific Research under award number FA9550-22-1-0312.
\end{acknowledgements}

\appendix
\section{\label{sec:level1}Electron-SPP matrix element}
\begin{figure}
  \centering
 \includegraphics[width=0.5\textwidth]{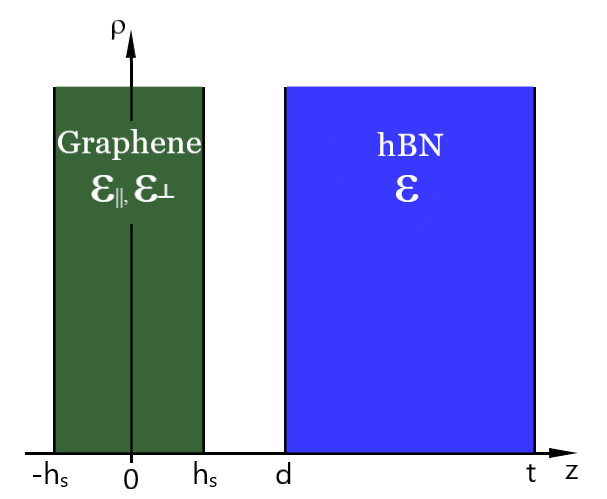}
 \caption{Schematic illustration of the cross-section of the graphene/hBN heterostructure. The monolayer or bilayer graphene is taken to be of thickness $2h_s$, with an anisotropic dielectric function $(\varepsilon_{\perp}, \varepsilon_{\parallel}(q))$ and centered around $z=0$. The hBN of thickness $t_{\rm hBN}=t-d$, and dielectric function $\varepsilon(\omega)$, supports the SPP. }
\label{Fig:Structure}
\end{figure}

To determine the electron-SPP coupling, we solve Maxwell’s equation for the spatial dependence of the electric potential $\varphi$ due to the SPP, using the following ansatz~\cite{fischetti2001effective}:
\begin{equation}
    \varphi(\boldsymbol{\rho},z,t)=\sum_{\boldsymbol{q}}\varphi(z)e^{i(\boldsymbol{q}.\boldsymbol{\rho}-\omega_{ph} t)}.
    \label{eq:anz}
\end{equation}
Here, $\boldsymbol{q}$ and $\boldsymbol{\rho}$ are the two-dimensional phonon wave vector and the spatial coordinate, respectively, and $\omega_{ph}$ is the phonon frequency. In isotropic materials, the Poisson equation $\nabla\varepsilon\nabla\varphi=0$ requires $\varphi(z)\propto e^{\pm q_zz}$ with $q_z=q$. However, in an anisotropic dielectric
$q_z=q\sqrt{\varepsilon_{\parallel}/\varepsilon_{\perp}}$.  Following Ref.~\cite{screenSPP_PRL2022}, we treat monolayer (bilayer) graphene as a dielectric layer of thickness $2h_s$, as shown in Fig.~\ref{Fig:Structure}, where $h_s=1.7$ \AA ($h_s=3.4$ \AA) for monolayer (bilayer) and accounts for the size of the electron cloud of the $\pi_z$ orbitals of carbon atoms. The thickness hBN $t_{\rm hBN}=t-d$ is placed at the van der Waals distance $d=3.4$ \AA ($d=5.1$ \AA) for monolayer (bilayer) graphene. 
Perpendicular to the plane, we choose the dielectric constant $\varepsilon_{\perp}=6$, as appropriate for the interface of bilayer graphene~\cite{bilayereps2019}. Within the plane, we choose a static dielectric function $\varepsilon_{\parallel}=1+v_c\Pi(q, E_F)$, where $v_c=2\pi e^2/q$, $\Pi(q, E_F)$ is the polarization function from the Random Phase Approximation, and we use the zero temperature limit~\cite{Hwang2008,DasSarma2010}.

The solution for $\varphi(z)$ in Eq.~(\ref{eq:anz}) has the form:
\begin{equation}\label{phiz}
      \varphi(z)=
      \begin{cases}
          A\;e^{q(z+h_s)}                    & z \leqslant -h_s,\\
          B\;e^{q_z(z-h_s)}+C\;e^{-q_z(z+h_s)}   & |z|<h_s,\\
          D\;e^{q(z-d)}+F\;e^{-q(z-h_s)}   & h_s\leqslant  z <d,\\
          E\;e^{q(z-t)}+G\;e^{-q(z-d)}   & d  \leqslant  z < t,\\
          H\;e^{-q(z-t)}                    & t \leqslant z.
      \end{cases}
\end{equation}
The coefficients $A$ $-$ $H$ and the dispersion relation for the SPPs are found using the boundary conditions $\varphi^+=\varphi^-$, $\varepsilon^+d\varphi^+/dz=\varepsilon^-d\varphi^-/dz$ at  $z=\pm h_s$ , $z=d$, and $z=t$, where the superscripts ``$+$'' and ``$-$'' indicate functions to the right and left of the boundaries, respectively.

The dielectric function hBN $\varepsilon(\omega)$ is given by:
\begin{equation}
  \varepsilon(\omega)=\frac{\epsilon_{\infty}\omega^2-\epsilon_{0}\omega_{TO}^2}{\omega^2-\omega_{TO}^2},
\end{equation}
where $\varepsilon_0=5.09$, $\varepsilon_{\infty}=4.575$, and $\hbar\omega_{TO}=97.3$ meV~\cite{Geick1966,perebeinos2010PRB}. (We omit the higher energy SPP branch at $\sim$200 meV.) The resulting solution for the SPP frequency is given by:
\begin{equation}
    \omega_{ph}(q)=\omega_{TO}\sqrt{\dfrac{\varepsilon_0+\alpha(q)}{\varepsilon_{\infty}+\alpha(q)}},
\label{Omega_SPP}
\end{equation}
where $\alpha(q)=-\varepsilon(\omega)$ is the solution of the dispersion relation.
To find the coefficients in Eq.~(\ref{phiz}) and the dispersion relation, we define $\varepsilon_{\rm ave} = \sqrt{\varepsilon_{\parallel}\varepsilon_{\perp}}$ and introduce the following variables:

\begin{equation}\label{Vars}
\begin{split}
  & C_d = {\rm cosh}(q (d-h_s)), \;\; T_d = {\rm tanh}(q (d-h_s)), \\
  & C_t = {\rm cosh}(q (t-d)), \;\; T_t = {\rm tanh}(q (t-d)), \\
  & C_z = {\rm cosh}(2q_zh_s), \;\; T_z = {\rm tanh}(2q_zh_s). 
\end{split}
\end{equation}

The coefficients are given by the following:
\begin{equation}\label{Coeff}
\begin{split}
    B =& A\frac{\varepsilon_{\rm ave}+1}{2\varepsilon_{\rm ave}}C_z(1+T_z),   \\
    C =& A\frac{\varepsilon_{\rm ave}-1}{2\varepsilon_{\rm ave}} ,  \\
    D =& \frac{B(\varepsilon_{\rm ave}+1)-C(\varepsilon_{\rm ave}-1)C_z(1-T_z)}{2}\\
    \times &C_d(1+T_d),   \\
    F =& \frac{B(1-\varepsilon_{\rm ave})+C(\varepsilon_{\rm ave}+1)C_Z(1-T_z)}{2},   \\
    E =& \frac{D(\varepsilon(\omega)+1)+F(\varepsilon(\omega)-1)C_d(1-T_d)}{2\varepsilon(\omega)}\\ 
    \times & C_t(1+T_t), \\
    G =& \frac{D(\varepsilon(\omega)-1)+F(\varepsilon(\omega)+1)C_d(1-T_d)}{2\varepsilon(\omega)}.
\end{split}
\end{equation}

The dispersion relation is obtained from the condition for the coefficient $H$: 
\begin{equation}\label{Dis}
\begin{split}
E+GC_t(1-T_t)=-\varepsilon(\omega)(E-GC_t(1-T_t)).
\end{split}
\end{equation}
After rearranging Eqs.~(\ref{Dis}) and (\ref{Coeff}), we find that $\varepsilon(\omega)=-\alpha(q)$, where $\alpha(q)$ is given by:
\begin{equation}\label{Alpha}
\begin{split}
    &\alpha(q) = \frac{b\pm\sqrt{b^2-4ac}}{2a}, \\
    a & = T_t(\varepsilon_{\rm ave} + T_z + T_d(\varepsilon_{\rm ave}+T_z\varepsilon^2_{\rm ave})),\\
    b & = (1+T_d)(2\varepsilon_{\rm ave} + T_z + T_z\varepsilon^2_{\rm ave}),\\
    c & = T_t(T_d(\varepsilon_{\rm ave} + T_z) + \varepsilon_{\rm ave}+T_z\varepsilon^2_{\rm ave}).
\end{split}
\end{equation}
The two different solutions of $\alpha(q)$ are due to the two surfaces of the finite thickness hBN. 

To find the form of the SPP potential that interacts with electrons in graphene \emph{i.e.}, $\varphi_0\equiv\varphi(z=0)=(B+C)e^{-q_zh_s}$ according to Eq.~(\ref{phiz}), we apply the normalization condition~\cite{paradisanos2020prominent,stroscio2001phonons}:
\begin{equation}\label{norm0}
    \dfrac{1}{L^2}\dfrac{\hbar}{2\omega}=\int\dfrac{1}{4\pi}\dfrac{1}{2\omega}(\dfrac{\partial\epsilon}{\partial\omega}|\mathbf{E}_{\perp}|^2+\dfrac{\partial\epsilon}{\partial\omega}|\mathbf{E}_{\parallel}|^2)dr,
\end{equation}
which allows us to solve for the magnitude of the coefficients $E^2+G^2$. In Eq.~(\ref{norm0}), $\mathbf{E}(\mathbf{r}) =-\nabla \varphi(\mathbf{r})$, $L^2=N_kA_c$ is the sample area, $A_c$ is the unit cell area, and $N_k$ is the number of k points. Finally, the e-SPP coupling constant $M_{\mathbf{k}\mathbf{q}}$ can be obtained as
\begin{equation}\label{eq:e-SPP}
\begin{split}
&    \vert M_{\mathbf{k}\mathbf{q}}\vert^2=(e\varphi_0)^2|\langle\psi_{\mathbf{k}}\vert\psi_{\mathbf{k+q}}\rangle|^2/N_k,  \\
 &   (e\varphi_0)^2=\dfrac{2\pi e^2}{qA_c}\hbar \omega \left(\dfrac{1}{\varepsilon_{\infty}+\alpha(q)}-\dfrac{1}{\varepsilon_{0}+\alpha(q)}\right)\\
&   \times  \frac{(B+C)^2}{E^2+G^2} \frac{C_z(1-T_z)(1+T_t)}{2T_t}. 
\end{split}
\end{equation}
Here, $\psi_{k}$ is a single particle wave function and the inner product in Eq.~(\ref{eq:e-SPP}) should be understood as corresponding to the wave function overlap in a primitive unit cell, not the entire sample. In the low-energy model, the wavefunction overlap between two states in the conduction band is  $|\langle\psi_{\mathbf{k}}\vert\psi_{\mathbf{k+q}}\rangle|^2=(1+\cos{(\theta_{kk+q})})/2$  for monolayer graphene, and $|\langle\psi_{\mathbf{k}}\vert\psi_{\mathbf{k+q}}\rangle|^2=(1+\cos{(2\theta_{kk+q})})/2$  for bilayer, where $\theta_{kk+q}$ is the angle between the two wavevectors $k$ and $k+q$~\cite{Novoselov2006}.

We note that the unscreened potential can be obtained by setting $h_s=0$ in the above equations and that the two SPP branches become degenerate in the absence of screening and infinitely thick hBN \emph{i.e}., $h_s=0$ and $t-d=\infty$. However, in this case, the electron-SPP coupling for each phonon branch would be half of the conventional coupling for semi-infinite  hBN~\cite{perebeinos2010PRB,Scharf_SPP_2013}. SPPs in the above solution form symmetric and antisymmetric linear combinations of two localized phonons at the two surfaces, with each of them contributing half of the coupling to electrons.  The screening of graphene breaks the symmetry and the SPP branch, which corresponds to the larger root $\alpha(q)$ in Eq.~(\ref{Alpha}) (or smaller SPP energy), gives the dominant electron-SPP coupling. At the same time, the smaller root of $\alpha(q)$ in Eq.~(\ref{Alpha}) gives negligible contribution at large values of $q$ and is consistently below 25\% throughout the range of values of $q$ and $t_{\rm hBN}$ reported in this study. 
To address that artifact of the model, which does not include losses in hBN, we use the larger root for $\alpha(q)$ in Eq.~(\ref{Alpha}) for the phonon energy, and contributions $(B+C)^2/(E^2+G^2)$ from both branches of SPP for the electron-SPP coupling. Note that in the limit of small $q$, $\alpha(q)=\infty$ and according to Eq.~(\ref{Omega_SPP}) $\omega_{ph}=\omega_{TO}$, whereas in the opposite limit $q=\infty$, $\alpha(q)=1$ and the conventional result for the SPP frequency corresponding to semi-infinite hBN and unscreened potential are obtained: $\omega_{ph}=\omega_{TO}\sqrt{(\varepsilon_0+1)/(\varepsilon_{\infty}+1)}$. Consequently, the overall SPP dispersion width, $\hbar\omega_{ph}( q=\infty)-\hbar\omega_{ph}(q=0)\approx4$ meV, is much smaller than the typical SPP energy $\hbar\omega_{ph}\approx 100$ meV.

\bibliography{reference.bib}

\end{document}